# Nonlinear terahertz Néel spin-orbit torques in antiferromagnetic Mn$_2$Au

Y. Behovits[1,2], A. L. Chekhov[1,2], S. Yu. Bodnar[3,4], O. Gueckstock[1,2], S. Reimers[3], T.S. Seifert[1,2], M. Wolf[2], O. Gomonay[3], M. Kläui[3], M. Jourdan[3] and T. Kampfrath[1,2*]

1. *Department of Physics, Freie Universität Berlin, 14195 Berlin, Germany*
2. *Department of Physical Chemistry, Fritz-Haber-Institut der Max-Planck-Gesellschaft, 14195 Berlin, Germany*
3. *Institute of Physics, Johannes Gutenberg-Universität Mainz, 55099 Mainz, Germany*
4. *Physikalisch-Chemisches Institut, Ruprecht-Karls-Universität Heidelberg, 69120 Heidelberg, Germany*

**Abstract**:

Antiferromagnets have large potential for ultrafast coherent switching of magnetic order with minimum heat dissipation. In novel materials such as Mn$_2$Au and CuMnAs, electric rather than magnetic fields may control antiferromagnetic order by Néel spin-orbit torques (NSOTs), which have, however, not been observed on ultrafast time scales yet. Here, we excite Mn$_2$Au thin films with phase-locked single-cycle terahertz electromagnetic pulses and monitor the spin response with femtosecond magneto-optic probes. We observe signals whose symmetry, dynamics, terahertz-field scaling and dependence on sample structure are fully consistent with a uniform in-plane antiferromagnetic magnon driven by field-like terahertz NSOTs with a torkance of $(150 \pm 50)\ \mathrm{cm}^2/\mathrm{A\ s}$. At incident terahertz electric fields above 500 kV/cm, we find pronounced nonlinear dynamics with massive Néel-vector deflections by as much as 30°. Our data are in excellent agreement with a micromagnetic model which indicates that fully coherent Néel-vector switching by 90° within 1 ps is within close reach.

# Figures

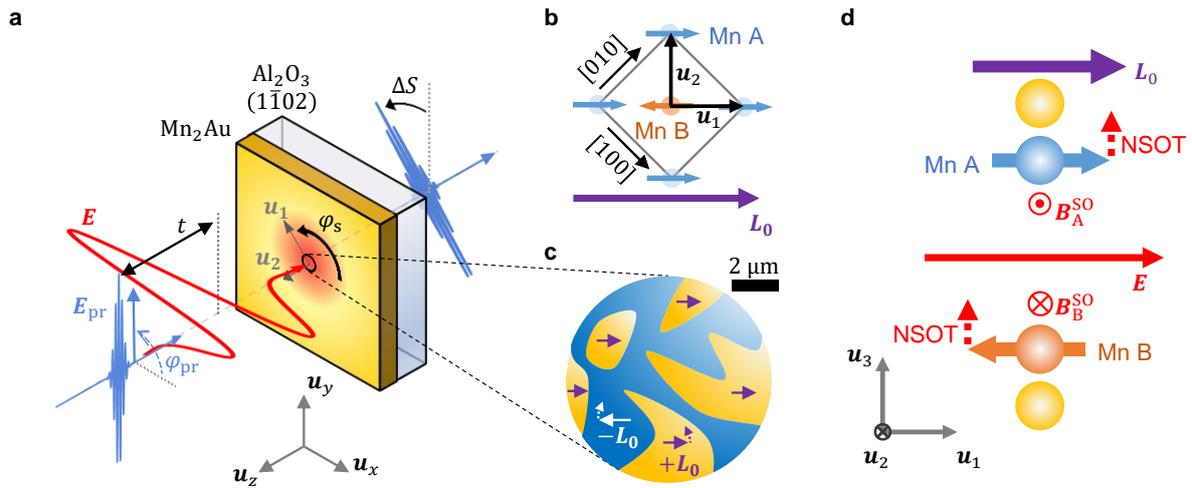

Fig. 1

**Fig. 1: Schematic of experiment and samples. (a)** A linearly polarized phase-locked terahertz pump pulse (red) with electric field $\boldsymbol{E} = E\boldsymbol{u}_x$ is normally incident onto an antiferromagnetic Mn₂Au thin film. The resulting spin dynamics are monitored by an optical probe pulse (blue) with field $\boldsymbol{E}_\mathrm{pr}$ by measuring its polarization change $\Delta S$ (rotation and ellipticity) vs delay time $t$ behind the sample. The probe-polarization angle $\varphi_\mathrm{pr} = \sphericalangle(\boldsymbol{E}_\mathrm{pr}, \boldsymbol{u}_x)$, sample azimuth $\varphi_\mathrm{s} = \sphericalangle(\boldsymbol{u}_1, \boldsymbol{u}_x)$ and terahertz field polarity ($\pm E$) can be varied in the laboratory frame $(\boldsymbol{u}_x, \boldsymbol{u}_y, \boldsymbol{u}_z)$. The sample-fixed frame is given by $\boldsymbol{u}_1 = [110]/\sqrt{2}$, $\boldsymbol{u}_2 = [\bar{1}10]/\sqrt{2}$ and $\boldsymbol{u}_3 = \boldsymbol{u}_z = [001]$. **(b)** Schematic of the Mn₂Au unit cell seen along $\boldsymbol{u}_z = \boldsymbol{u}_3$. The local magnetic moments on the Mn A and Mn B sites lead to the sublattice magnetizations $\boldsymbol{M}_\mathrm{A}$ and $\boldsymbol{M}_\mathrm{B}$, respectively. The Néel vector $\boldsymbol{L} = \boldsymbol{M}_\mathrm{A} - \boldsymbol{M}_\mathrm{B}$ (purple arrow) equals $\boldsymbol{L}_0$ before pump excitation. For simplicity, the Au atoms are omitted. **(c)** In the magnetically prealigned sample, the in-plane distribution $\boldsymbol{L}_0(x,y)$ consists of regions with $\boldsymbol{L}_0 \upuparrows +\boldsymbol{u}_1$ and $\boldsymbol{L}_0 \upuparrows -\boldsymbol{u}_1$, resulting in a pattern of 180° domains. Dashed arrows indicate a possible pump-induced deflection $\Delta\boldsymbol{L}$. **(d)** An in-plane electric field $\boldsymbol{E}$ induces a staggered spin-orbit fields $\boldsymbol{B}_\mathrm{A}^\mathrm{SO} = -\boldsymbol{B}_\mathrm{B}^\mathrm{SO}$. For $\boldsymbol{E} \parallel \boldsymbol{L}_0$, the resulting NSOTs on Mn A and Mn B moments are maximum and directed out-of-plane ($\parallel \boldsymbol{u}_3$).

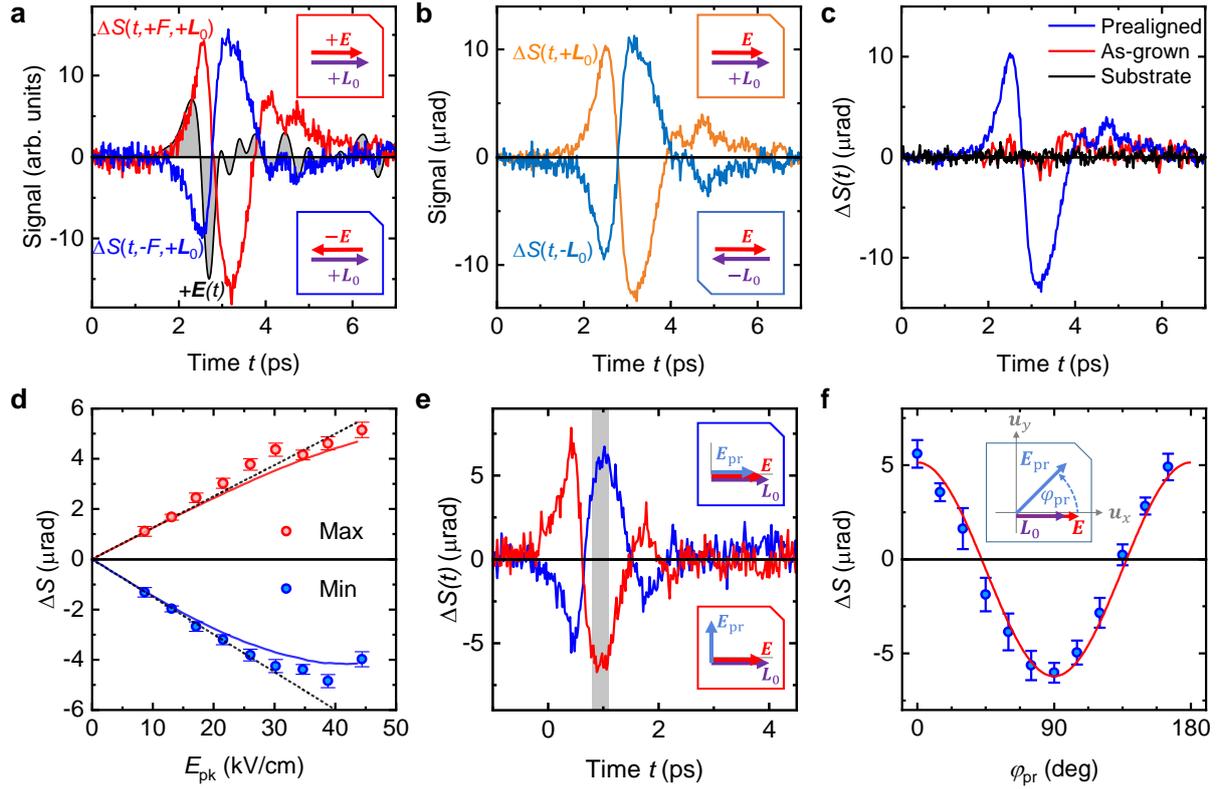

Fig. 2

**Fig. 2: Signatures of terahertz spin dynamics. (a)** Pump-induced probe polarization rotation $\Delta S(t, +F, L_0)$ (red) and $\Delta S(t, -F, L_0)$ (blue) vs pump-probe delay $t$ measured for opposite polarities of the terahertz pump field $F = (E, B)$ in the prealigned sample with $\varphi_s = \varphi_{pr} = 0°$. The grey-shaded area shows the terahertz electric field $E(t)$ for reference. The incident peak terahertz field in air is 250 kV/cm, corresponding to $E_{pk} = 15$ kV/cm inside the sample. **(b)** Signal components $\Delta S(t, +L_0)$ ($\varphi_s = 0°$, orange solid line) and $\Delta S(t, -L_0)$ ($\varphi_s = 180°$, blue) odd in the driving terahertz field (Eq. (1)). **(c)** Signal component $\Delta S(t) = [\Delta S(t, +L_0) - \Delta S(t, -L_0)]/2$ odd in both driving field $F$ and Néel vector $L_0$ for prealigned (blue line) and as-grown (red) sample. The black line shows the signal from the bare Al$_2$O$_3$ substrate. **(d)** Maximum (red dots) and minimum value (blue dots) of signal $\Delta S(t)$ odd in $F$ and $L_0$ vs peak terahertz electric field $E_{pk}$ inside the sample. Data was taken from a different sample region than in panels (a)-(c), yielding smaller signal magnitudes. The red and blue solid line is a model calculation (Fig. 4 and Eq. (16)). The dotted black lines indicate the linear approximation of the model. **(e)** Odd-in-$F$ signal $\Delta S(t, +L_0)$ for $E_{pk} = 15$ kV/cm and $\varphi_s = \varphi_{pr} = 0°$ (blue) and 90° (red). The inset indicates the probe polarization in the lab frame. For these measurements, a second Al$_2$O$_3$ substrate was used to compensate the birefringence of the first substrate. **(f)** Time-average of signal over gray area in panel (e) vs incident probe polarization angle $\varphi_{pr}$ (blue circles) and reference curve $\cos(2\varphi_{pr})$. The inset indicates the probe polarization rotation in the lab frame. Signals can be found in Fig. S8a.

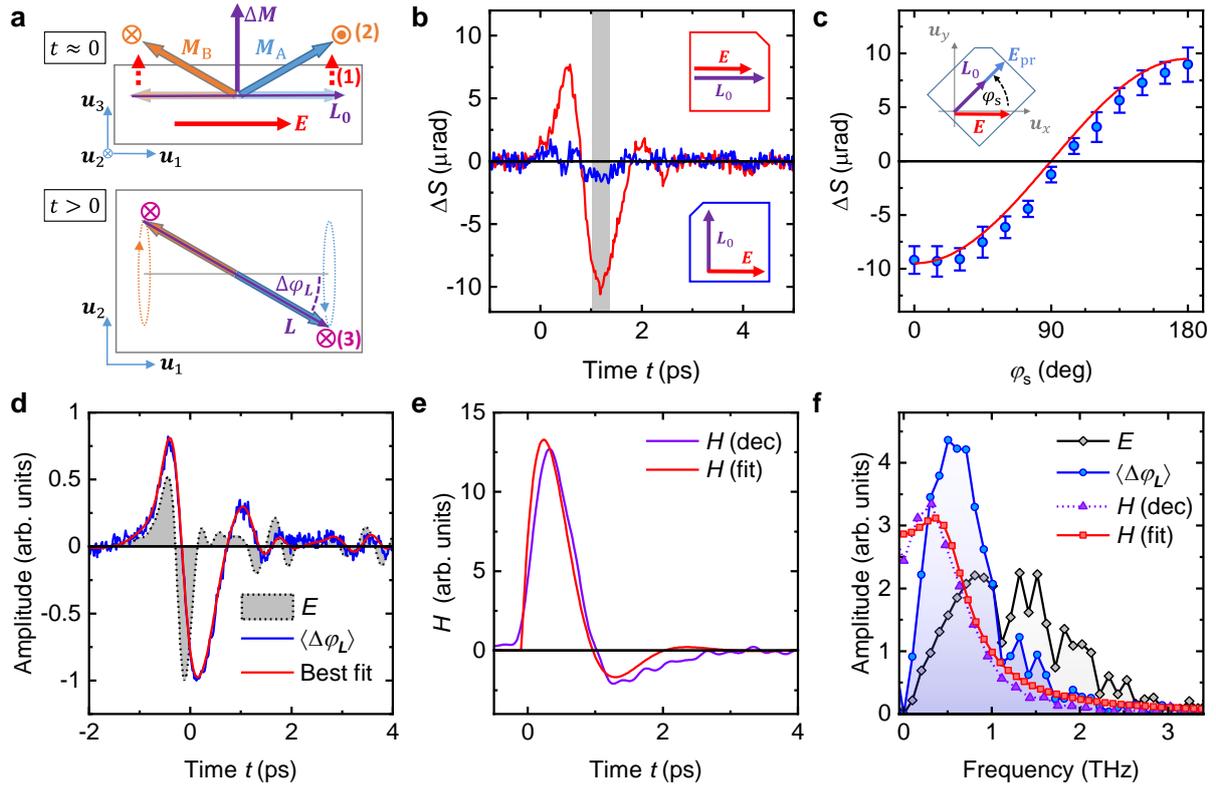

Fig. 3

**Fig. 3: Terahertz antiferromagnetic magnon driven by NSOTs. (a)** Expected dynamics driven by terahertz NSOTs, showing Mn sublattice spins A and B (orange and blue arrows). In step (1), an impulsive electric field $E(t) = A\delta(t)$ with strength $A$ induces staggered spin-orbit fields $\propto \pm u_z \times A$ at A, B. Owing to $M_{A0} = -M_{B0} = L_0/2$, the field-like torques (red-dashed lines) on A, B are equal and $\propto L_0 \times (u_z \times E)$. They cant $M_A$, $M_B$ and induce an out-of-plane magnetization $\Delta M \propto L_0 \times (u_z \times A)$ at time $t = 0^+$. (2) Subsequently, the exchange field $B_{ex}L_0/|L_0|$ gradually deflects the Néel vector by $\Delta L \propto (L_0/|L_0|) \times \Delta M$ (orange symbols). (3) The anisotropy fields induce torques $\parallel u_3$ (magenta symbols) and, thus, precession of $M_A$ and $M_B$ (dotted lines). The resulting dynamic deflection $\Delta L \propto \Delta \varphi_L$ lies within the plane $\perp u_3$ and fulfills $|\Delta L| \gg |\Delta M|$. **(b)** $\Delta S(t) \propto \langle \Delta \varphi_L(t) \rangle$ for $\varphi_s = 0°$ (red) and 90° (blue) with $E_{pk} = 15$ kV/cm. The probe polarization is rotated together with the sample such that $\varphi_{pr} = \varphi_s$. The insets show the relative orientation of $E$ and $L_0$. **(c)** Time-average of signal over the gray area in panel (c) vs $\varphi_s$ (blue circles), along with a $\cos \varphi_s$ curve (red line) that is expected from NSOTs. The inset schematically shows the simultaneous rotation of sample and probe polarization to maintain $\varphi_{pr} = \varphi_s$. Signals can be found in Fig. S8b. **(d)** Volume-averaged in-plane deflection angle $\langle \Delta \varphi_L(t) \rangle \propto \Delta S(t)$ of the Néel vector (blue) and the driving terahertz electric field $E(t)$ (gray-shaded area) with $E_{pk} = 15$ kV/cm. The red curve is the best fit of Eq. (4) to $\Delta S(t)$. **(e)** Time-domain impulse response $H(t)$ of $\Delta \varphi_L$ as obtained from fitting (panel (d), red line) or by model-free deconvolution of Eq. (4) (purple line). **(f)** Fourier amplitude spectra of $E(t)$ (gray diamonds) and $\langle \Delta \varphi_L(t) \rangle$ (blue circles) of panel (d) and of the extracted impulse response functions $H(t)$ of panel (e) (red squares, purple triangles).

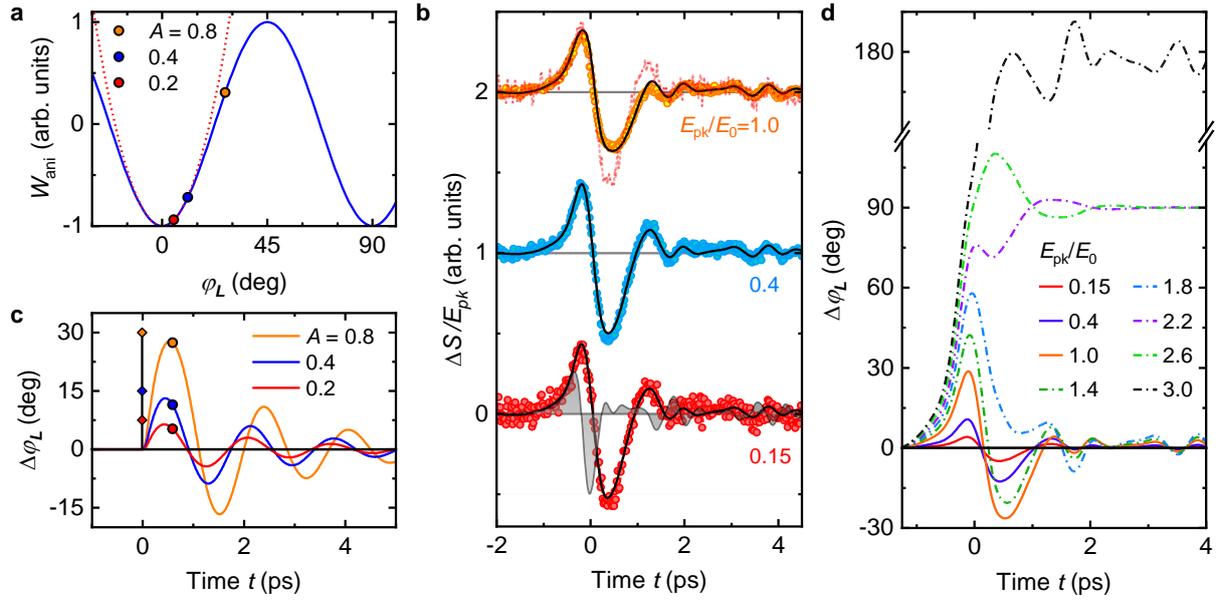

Fig. 4

**Fig. 4: Nonlinear Néel-vector precession and extrapolation to switching.** (a) Magnetic anisotropy energy $W_{\text{ani}} \propto -B_{\text{ani}} B_{\text{ex}} \cos(4\varphi_L)$ vs azimuthal rotation $\varphi_L$ of the Néel vector (blue solid line) and its harmonic approximation (red dotted line). The red, blue and orange dot indicates, respectively, the position and anisotropy energy at $t = 0.6$ ps following fictitious impulsive excitation by $E(t) \propto A\delta(t)$ with relative strength $A = 0.2$, $0.4$ and $0.8$. (b) Normalized measured signals $\Delta S(t)/E_{\text{pk}}$ for normalized terahertz peak fields $E_{\text{pk}}/E_0$, where $E_0 = 40$ kV/cm is the maximum available field inside the sample. Signals are vertically offset for clarity. The gray-shaded area shows the incident driving field $E(t)$ for reference. Black lines are a joint fit using the model of Eq. (16). The red dotted line shows the signal for $E_{\text{pk}}/E_0 = 0.15$ for comparison. (c) Calculated dynamics $\Delta\varphi_L(t)$ for the three excitation amplitudes $A$ (diamonds) in panel (a). Dots correspond to $\Delta\varphi_L(t)$ at $t = 0.6$ ps (see panel (a)). For illustrative purposes, the Gilbert damping parameter is chosen smaller than the experimental value. (d) Simulated dynamics of the deflection angle $\Delta\varphi_L$ for various driving fields $E_{\text{pk}}/E_0$. Solid lines correspond to the fits from panel (b) ($E_{\text{pk}}/E_0 \leq 1$), whereas dash-dotted lines are simulations for moderately larger fields ($E_{\text{pk}}/E_0 > 1$).

**Introduction.** Antiferromagnets offer great potential for robust, ultrafast and space- and energy-efficient spintronic functionalities [1-6]. Remarkably, in a recently discovered class of metallic antiferromagnets with locally broken inversion symmetry, coherent rotation of the Néel vector $L$ should be possible by simple application of electrical currents. This fascinating phenomenon is driven by the Néel spin-orbit torques (NSOTs) [7-9] that arise from staggered spin-orbit fields at the antiferromagnetically coupled spin sublattices [10, 11]. So far, switching studies of CuMnAs and $Mn_2Au$ showed indications of NSOTs [12-15], but also a strong and possibly dominant heat-driven reorientation of $L$ [16, 17].

To reveal direct signatures of NSOTs and gauge their potential for ultrafast coherent antiferromagnetic switching, electric fields at terahertz frequencies are particularly interesting because they are often resonant with long-wavelength antiferromagnetic magnons [18-23]. Consequently, terahertz NSOTs should require significantly smaller current amplitudes to modify antiferromagnetic order and, consequently, mitigate unwanted effects such as Joule heating. Studying terahertz NSOTs should also provide fundamental insights into key parameters such as torkance and the frequency and lifetime of long-wavelength magnons in the novel antiferromagnets.

Probing antiferromagnetic responses is highly nontrivial because, unlike the magnetization of ferromagnets, the Néel vector $L$ cannot be controlled by external magnetic fields below several tesla in most compounds [24-26]. Consequently, experimental separation of magnetic and nonmagnetic dynamical effects is challenging [27, 28]. Moreover, the small size of antiferromagnetic domains [24, 29] implies that the spatial average of the Néel vector and the NSOTs vanishes.

In this work, we report on terahertz-pump magneto-optic-probe experiments on $Mn_2Au$ thin films (Fig. 1a). We find birefringence signals linear in the incident terahertz electric-field transient. They can consistently be assigned to a uniform, strongly damped and coherent antiferromagnetic magnon at 0.6 THz that is excited by field-like NSOTs. When the terahertz field inside the $Mn_2Au$ film exceeds 30 kV/cm, the Néel-vector dynamics become significantly nonlinear. Comparison to an analytical model allows us to extract values of NSOT torkance, magnon frequency and Gilbert damping, all of which are consistent with previous predictions and experiments. We deduce that the Néel vector $L$ is transiently deflected by as much as 30° at the maximum peak field of 40 kV/cm.

Our results imply that terahertz electric fields and NSOTs can drive coherent nonlinear magnon dynamics in $Mn_2Au$, thereby bringing coherent switching, a central goal of antiferromagnetic spintronics, in close reach. Indeed, extrapolation of our data indicates that coherent rotation of $L$ by 90° can be achieved by terahertz pulses with a moderately increased peak field strength of around 120 kV/cm inside $Mn_2Au$.

**Experiment.** We excite $Mn_2Au$ thin films with intense phase-locked single-cycle terahertz pulses to drive spin dynamics and magneto-optically monitor them with a femtosecond probe pulse (Fig. 1a).

$Mn_2Au$ is a metallic collinear antiferromagnet with high Néel temperature (>1000 K). It has two spin sublattices A and B with magnetization $M_A$ and $M_B$, resulting in net magnetization $M = M_A + M_B$ and the Néel vector $L = M_A - M_B$ (Fig. 1b). In equilibrium, one has $M = M_0 = 0$ and $L = L_0$. $Mn_2Au$ is an easy-plane antiferromagnet with strong out-of-plane ($u_3 = [001]$) and a 1-2 orders of magnitude smaller biaxial in-plane anisotropy field [24, 30-32]. Consequently, the vectors $M_A$ and $M_B$ align along one of the easy axes $u_1 = [110]/\sqrt{2}$ or $u_2 = [\bar{1}10]/\sqrt{2}$ (Fig. 1b). Due to locally broken inversion symmetry and strong spin-orbit coupling, an electric current leads to staggered spin-orbit fields at the A and B sites, which exert NSOTs on the localized magnetic moments (Fig. 1d) [11].

We study two epitaxially grown $Mn_2Au$(001) thin films (thickness of 50 nm) with an $Al_2O_3$ cap layer (3 nm) on $Al_2O_3(1\bar{1}02)$ substrates (500 μm). In the as-grown samples, the four equilibrium directions of the Néel vector $L_0$ are at 0°, 90°, 180° and 270° relative to $u_1$ [24, 29]. The resulting domains have a size of the order of 1 μm with approximately equal distribution over the film plane and a small strain-induced preference along one easy axis [33]. Following growth, one sample was subject to an intense magnetic-field pulse [24, 34, 35], which aligned most domains at 0° or 180° (Fig. 1c). For test purposes, we also consider $Mn_2Au$|Py samples in which the volume-averaged Neel vector $\langle L_0 \rangle$ is oriented parallel to the magnetization of the exchange-coupled ferromagnetic Py (permalloy $Ni_{80}Fe_{20}$) layer [29].

In our ultrafast setup (Figs. 1a, S1, S2), the terahertz pump and optical probe beam are both normally incident onto the sample. The driving terahertz field $F = (E, B)$ contains the electric component $E = Eu_x$, shown in Fig. 2a, and the magnetic component $\boldsymbol{B} = B\boldsymbol{u}_y$, which exhibits the same shape. In free space, $E$ reaches peak values up to 600 kV/cm, which is reduced to 6% inside the Mn₂Au film. The resulting spin dynamics are monitored by linearly polarized optical probe pulses [36] with the polarization plane at an angle $\varphi_{\mathrm{pr}}$ relative to the laboratory axis $\boldsymbol{u}_x$ (Fig. 1a). The detected signals $\Delta S(t, F, \boldsymbol{L}_0)$ are the pump-induced probe-polarization rotation and ellipticity vs pump-probe delay $t$. The sign of the pump field ($\pm F$) and the local Néel vector ($\pm \boldsymbol{L}_0$) can be reversed by, respectively, a pair of polarizers and rotation of the sample azimuth $\varphi_{\mathrm{s}} = \sphericalangle(\boldsymbol{u}_1, \boldsymbol{u}_x)$ by 180°.

**Raw data.** Excitation with terahertz field $+F(t)$ results in the pump-induced rotation signal $\Delta S(t, +F, \boldsymbol{L}_0)$ (red curve in Fig. 2a), as measured for the magnetically prealigned sample with $\varphi_{\mathrm{pr}} = \varphi_{\mathrm{s}} = 0°$ (Fig. 1a). When the terahertz field is reversed, the induced signal $\Delta S(t, -F, \boldsymbol{L}_0)$ (blue curve in Fig. 2a) changes sign, too, as expected for NSOTs. Therefore, we consider the signal component that is odd with respect to $F$, i.e.,

$$\Delta S(t, \boldsymbol{L}_0) = \frac{\Delta S(t, +F, \boldsymbol{L}_0) - \Delta S(t, -F, \boldsymbol{L}_0)}{2}. \quad (1)$$

Fig. 2b displays $\Delta S(t, \pm \boldsymbol{L}_0)$ for opposite local Néel vectors. Again, the two signals are approximately reversed versions of each other, pointing to a strong contribution of the antiferromagnetic order. Analogous to Eq. (1), we determine signals $\Delta S(t)$ odd in both the pump field $F$ and the Néel vector $\boldsymbol{L}_0$ and focus on them in the following.

To check further that $\Delta S(t)$ reports on magnetic order, Fig. 2c displays waveforms $\Delta S(t)$ from the prealigned and as-grown Mn₂Au film. Remarkably, the signal from the prealigned sample is more than a factor of 5 larger and independent of the probed spot over areas much larger than the individual antiferromagnetic domains. In contrast, $\Delta S(t)$ from the as-grown sample is typically within the experimental noise floor (see Fig. S5) and only exceptionally large for the example waveform depicted in Fig. 2c. For the bare Al₂O₃ substrate, the signal odd in $F$ is zero within the experimental accuracy. The different response of the as-grown and prealigned sample (Fig. 2c) and the very similar response of the prealigned and Mn₂Au|Py test sample (Supplementary Note 11) provide strong evidence that the component $\Delta S(t)$ is related to the antiferromagnetic order of Mn₂Au.

Fig. 2d shows that the absolute maximum of $\Delta S(t)$ grows roughly linearly with the peak amplitude $E_{\mathrm{pk}}$ of the terahertz electric field inside the sample over the full range of $E_{\mathrm{pk}}$. In contrast, the absolute minimum of $\Delta S$ features an onset of nonlinear behavior for $E_{\mathrm{pk}} > 30$ kV/cm (see black-dotted lines and Fig. S3). Therefore, the signals in Fig. 2b are the linear response to the driving terahertz field ($|E| < 15$ kV/cm), and we will first focus on this lowest-order perturbation regime.

**Signal phenomenology.** To understand the pump-probe signal, we need to relate it to the perturbing terahertz pump field $F = (\boldsymbol{E}, \boldsymbol{B})$ and the instantaneous magnetic order, which is quantified by $\boldsymbol{L}$ and $\boldsymbol{M}$. The symmetry properties of Mn₂Au impose strict conditions on the relationship between $\Delta S$ and $F$, $\boldsymbol{L}$, $\boldsymbol{M}$, as detailed in the Methods. In particular, we find that any signal linear in $F$ and odd in $\boldsymbol{L}_0$ (Fig. 2c) arises entirely from the terahertz electric field $\boldsymbol{E}$ and not the magnetic field $\boldsymbol{B}$.

Generally, the signal depends on pump-induced changes in $\boldsymbol{L}$, $\boldsymbol{M}$, and the non-spin degrees of freedom. Our experimental geometry (Fig. 1a) and the point-symmetry group of Mn₂Au [31] imply that the pump-induced signal is up to second order in $\boldsymbol{L}$ and $\boldsymbol{M}$ given by the form

$$\Delta S \propto a \sin(2\varphi_{\mathrm{pr}} - 2\varphi_{\mathrm{s}}) \Delta \langle L_\parallel^2 \cos(2\varphi_L) \rangle + b \cos(2\varphi_{\mathrm{pr}} - 2\varphi_{\mathrm{s}}) \Delta \langle L_\parallel^2 \sin(2\varphi_L) \rangle + c\boldsymbol{u}_z \cdot \langle \Delta \boldsymbol{M} \rangle + \Delta S_\mathcal{N}. \quad (2)$$

Here, $a$, $b$ and $c$ are sample- and setup-dependent coefficients, $\varphi_{\mathrm{pr}} - \varphi_{\mathrm{s}} = \sphericalangle(\boldsymbol{E}_{\mathrm{pr}}, \boldsymbol{u}_1)$ (Fig. 1a), and $\boldsymbol{L}_\parallel = \boldsymbol{L} - (\boldsymbol{u}_z \cdot \boldsymbol{L})\boldsymbol{u}_z$ is the in-plane projection of the Néel vector with azimuthal angle $\varphi_L = \sphericalangle(\boldsymbol{L}_\parallel, \boldsymbol{u}_1)$. A preceding $\Delta$ denotes pump-induced changes, and $\langle . \rangle$ means spatial averaging over the probed Mn₂Au volume. The first and second term of Eq. (2) are quadratic in $\boldsymbol{L}_\parallel$ (magnetic linear birefringence (MLB) [28, 33]) and monitor spin dynamics through changes in $L_\parallel^2$ and $\varphi_L$. The third term is linear in $\boldsymbol{M}$ (magnetic circular birefringence (MCB) [28]) and reports on out-of-plane variations of $\boldsymbol{M}$.

Note that the last term $\Delta S_\mathcal{N}$ is unrelated to transient changes in $\boldsymbol{L}$ and $\boldsymbol{M}$. It exclusively arises from variations of the non-spin degrees of freedom $\mathcal{N}$, such as phonons, and can be shown to exhibit the same dependence on $\varphi_E$, $\varphi_{L_0}$ and $\varphi_s$ as the first three terms (see Methods). Nonmagnetic contributions of this kind are rarely considered [37]. Figs. S4 and S10, however, show that $\Delta S_\mathcal{N}$ does not make a dominant contribution to our total signal $\Delta S(t)$.

To identify the dominant terms in Eq. (2), we vary the incoming probe polarization $\varphi_{\text{pr}}$ while keeping $\varphi_s = 0°$ (Fig. 2e). We find that the waveforms $\Delta S(t, \boldsymbol{L}_0)$ exhibit opposite sign for $\varphi_{\text{pr}} = 0°$ and $90°$ (Fig. 2e). Their amplitude follows a $\cos(2\varphi_{\text{pr}})$ dependence to very good approximation (Fig. 2f). Therefore, the signal predominantly derives from the second term of Eq. (2). Assuming the pump-induced changes in $L_\parallel^2$ and $\varphi_L$ are small, we linearize this term. In the prealigned sample, we have $\sin(2\varphi_{L_0}) = 0$ for both the 0° and 180° domains, and Eq. (2) simplifies to

$$\Delta S(t) \propto L_{\parallel 0}^2 \langle \Delta \varphi_L(t) \rangle. \tag{3}$$

This important result implies that the signal $\Delta S$ directly monitors the dynamics of the spatially averaged azimuthal rotation $\Delta \varphi_L$ of $\boldsymbol{L}$, with the rotation angle being odd in $\boldsymbol{L}_0$ and odd in $\boldsymbol{E}$.

Note that a nonzero $\langle \Delta \varphi_L(t) \rangle$ requires a non-vanishing average Neel vector $\langle \boldsymbol{L}_0 \rangle$ in the probed volume. We can understand the occurrence of $\langle \boldsymbol{L}_0 \rangle \neq 0$ by the magnetic prealignment procedure (see Methods and Supplementary Note 5).

**NSOT-driven terahertz magnon.** As the signal $\Delta S(t)$ cannot arise from the terahertz magnetic field, we can discard Zeeman torque ($\propto \boldsymbol{B}$) and field-derivative torque ($\propto \partial \boldsymbol{B}/\partial t$) [38, 39] as possible microscopic mechanisms. Effects of Joule heating are excluded, too, as they would scale quadratically with the pump field.

Because the Néel-vector rotation $\Delta \varphi_L$ is linear in the terahertz electric field $\boldsymbol{E}$, it may arise from bulk NSOTs. To put this conjecture to the test, we consider the effect of NSOTs on the magnetic order of Mn$_2$Au. As detailed in Fig. 3a for a single $\boldsymbol{L}_0$ domain, NSOTs induce an in-plane deflection $\Delta \varphi_L \propto \boldsymbol{u}_z \cdot (\boldsymbol{L}_0 \times \Delta \boldsymbol{L})$ of the Néel vector. As the NSOT-induced change $\Delta \boldsymbol{L}$ is even in $\boldsymbol{L}_0$ (Fig. 3a), $\Delta \varphi_L$ is odd in $\boldsymbol{L}_0$, precisely as $\Delta S(t)$ (Eq. (3)). The amplitude of $\Delta \varphi_L$ is proportional to $\cos \sphericalangle (\boldsymbol{E}, \boldsymbol{L}_0)$. To test whether the measured $\Delta S(t)$ follows the same dependence, we rotate the sample and, thus, the Néel vector (Fig. 1a). We find that $\Delta S(t)$ indeed scales with $\cos \sphericalangle (\boldsymbol{E}, \boldsymbol{L}_0)$, as shown in Fig. 3b, c.

To summarize, the signal $\Delta S(t) \propto \langle \Delta \varphi_L(t) \rangle$ is fully consistent with the terahertz-NSOTs scenario of Fig. 3a owing to its phenomenology: linear in $\boldsymbol{E}$, odd in $\boldsymbol{L}_0$ and scaling with $\cos \sphericalangle (\boldsymbol{E}, \boldsymbol{L}_0)$. The resulting predominant in-plane motion of $\boldsymbol{L} = \boldsymbol{L}_0 + \Delta \boldsymbol{L}$ corresponds to the in-plane magnon mode of Mn$_2$Au. It is accompanied by an out-of-plane magnetization $|\Delta \boldsymbol{M}| \propto |\partial \Delta \boldsymbol{L}/\partial t|$ (Fig. 3a) [40, 41], which is more than 2 orders of magnitude smaller than $\Delta \boldsymbol{L}$ and below our detection sensitivity. The remaining second magnon mode of Mn$_2$Au [32, 40], which involves an out-of-plane oscillation of $\boldsymbol{L}$, would be even in $\boldsymbol{L}_0$ and possibly be masked by contributions independent of magnetic order (see Methods).

**Micromagnetic model.** From Fig. 3a, we expect a harmonic time-dependence of $\Delta \varphi_L$ that starts sine-like. Indeed, as detailed in the Methods, the rotation $\Delta \varphi_L$ can be described as deflection of a damped oscillator [40, 42] whose potential energy $W_{\text{ani}} \propto -B_{\text{ani}} B_{\text{ex}} \cos(4\varphi_L)$ (Fig. 4a) is determined by the in-plane anisotropy field $B_{\text{ani}}$ and exchange field $B_{\text{ex}}$ of the Mn$_2$Au thin film. The damping of the oscillator is proportional to $B_{\text{ex}}$ and the Gilbert parameter $\alpha_G$. The driving force scales with $\lambda_{\text{NSOT}} \sigma E(t)$, where $\lambda_{\text{NSOT}}$ is the NSOT coupling strength (or torkance), and $\sigma \approx 1.5 \text{ MS/m}$ is the measured terahertz conductivity (Fig. S6).

At times $t < 0$, we have equilibrium with $\varphi_L = 0°$, 90°, 180° or 270° (Fig. 4a). Following a fictitious impulsive electric field $A\delta(t)$ with small amplitude $A$, the response $\Delta \varphi_L$ is linear and yields a damped sinusoidal oscillation $H(t) \propto \Theta(t) e^{-\Gamma t} \sin(\Omega t)$ with the Heaviside step function $\Theta(t)$, frequency $\Omega/2\pi = \sqrt{\Omega_0^2 - \Gamma^2}/2\pi$, bare resonance frequency $\Omega_0/2\pi$ and damping rate $\Gamma$ [40]. For an arbitrary driving field $E(t)$, the solution of the linearized Eq. (16) is given by the superposition (convolution)

$$\Delta \varphi_L(t) = (H * E)(t) = \int d\tau\, E(\tau) H(t - \tau). \tag{4}$$

Fig. 3d displays the measured $E(t)$ and $\Delta S(t) \propto \langle \Delta\varphi_L(t)\rangle$ along with a fit by Eq. (4), where $\Omega_0$, $\Gamma$ and a global amplitude factor are free parameters. Eq. (4) provides an excellent description for $\Omega_0/2\pi = (0.6 \pm 0.1)$ THz and $\Gamma/2\pi = (0.30 \pm 0.05)$ THz. The resulting modeled impulse response function $H(t)$ (Fig. 3e) is a strongly damped oscillation that decays within about two oscillation cycles.

As a cross-check, we solve Eq. (4) for $H(t)$ by numerical deconvolution without model assumptions. We find that the deconvoluted response agrees well with the fit-based result in both the time (Fig. 3e) and frequency domain (Fig. 3f). The amplitude spectrum of the transient deflection $\Delta\varphi_L$ illustrates the broad resonance-like response given by $H$ (Fig. 3f).

Our experimentally obtained magnon frequency lies outside the accessible range of previous Brillouin- and Raman-scattering measurements [32]. A peak at 0.12 THz was ascribed to the in-plane magnon mode, but the magnetic origin of this feature is not confirmed. The bare magnon frequency $\Omega_0/2\pi$ allows us to estimate the spin-flop field by $B_{sf} \sim \Omega_0/\gamma$, where $\gamma$ is the gyromagnetic ratio (see Methods). We infer $B_{sf} \sim 20$ T at a temperature of 300 K, which is consistent with a previous order-of-magnitude estimate of 30 T at 4 K [24, 43]. A more accurate comparison requires a detailed understanding of the magnetic reordering processes at high magnetic fields, which remain elusive. With $B_{ex} = 1300$ T [31], we extract a Gilbert-damping parameter of $\alpha_G = \Gamma/\gamma B_{ex} = 0.008$, which is consistent with theoretical predictions for metallic antiferromagnets [44, 45] and recent studies of optically driven spin dynamics in IrMn [46].

**Nonlinear regime and torkance.** Finally, we increase the peak terahertz field $E_{pk}$ inside the sample above 30 kV/cm to study the nonlinear response of Mn$_2$Au indicated by Fig. 2d. Examples of signal waveforms $\Delta S(t)$, normalized to $E_{pk}$, are shown in Fig. 4b. Remarkably, as $E_{pk}$ grows (red to orange curve), the normalized signal not only decreases its peak value but also becomes more symmetric.

Qualitatively, this transition from a linear to nonlinear response can be well understood by the anharmonic potential of Fig. 4a. As $W_{ani}$ grows sub-quadratically for large deflection angles $\Delta\varphi_L$, smaller restoring forces and, thus, slower dynamics than in the harmonic case result. Importantly, in the anharmonic regime, the temporal waveform is unambiguously connected to the excitation strength, as illustrated for the fictitious impulsive driving forces in Fig. 4c. Further, as the magneto-optic signal is governed by the second term of Eq. (2), it exhibits a sub-linear growth at large deflection angles $\Delta\varphi_L$.

Quantitatively, we fit the measured signals $\Delta S(t) \propto \langle\sin[2\Delta\varphi_L(t)]\rangle$ (Fig. 4b) by numerically solving the micromagnetic model (see Methods), where a scaling factor and the torkance $\lambda_{NSOT}$ (Eq. (16)) are the only free fit parameters. As the dynamics of the 0° and 180° domains are given by $\pm\Delta\varphi_L(t)$, the spatial average $\Delta S(t)$ reflects the dynamics of a single domain, but just with decreased signal amplitude. The inferred $\lambda_{NSOT} = (150 \pm 50)$ cm$^2$/A s is, to our knowledge, a first experimental determination of an NSOT torkance. It corresponds to a staggered field of $(8 \pm 3)$ mT per $10^7$ A/cm$^2$ driving current density, which agrees well with ab initio calculations that found 2 mT per $10^7$ A/cm$^2$ [11]. Our procedure also allows us to calculate the signal amplitudes vs the terahertz peak field $E_{pk}$. As seen in Fig. 2d, the onset of non-linearity is in good agreement with our experiment.

The calculated Néel-vector dynamics $\Delta\varphi_L(t)$ of a single domain are shown in Fig. 4d. We find that the rotation angle $\Delta\varphi_L$ reaches 30° at the maximum terahertz peak field of 40 kV/cm inside the sample. This massive deflection is about 2 orders of magnitude larger than the magnetization deflection in ferromagnets that were induced by incident terahertz pulses comparable to ours [47, 48]. It illustrates the benefits of NSOTs and exchange-enhancement of the resonant Néel vector response in antiferromagnets.

Our modeling enables us to extrapolate the dynamics to even higher driving fields (Fig. 4d). Remarkably, at a peak amplitude exceeding our maximum available incident field of 600 kV/cm by only a factor of 2.5, the Néel vector overcomes the potential-energy maximum of the magnetic anisotropy at $\varphi_L = 45°$ (Fig. 4a) and coherently switches from $\varphi_L = 0°$ to $90°$. The ultrafast switching time of only 1 ps is given by half the period $\pi/\Omega$ of the terahertz magnon. We estimate that the temperature increase due to terahertz pulse absorption is less than 5 K (Supplementary Note 9). Therefore, resonant NSOTs induced by terahertz electric pulses are an ideal driver to achieve coherent ultrafast and energy-efficient antiferromagnetic switching. When the terahertz field is increased by a factor of 3.0, we even obtain switching to $\varphi_L = 180°$ (Fig. 4d).

**Conclusion.** It is shown that terahertz electric fields exert field-like NSOTs in $Mn_2Au$ antiferromagnetic thin films. Using magnetic linear birefringence, we observe a strongly damped precession of the Néel vector in the sample plane at 0.6 THz. Our interpretation is consistent with regard to the symmetry and dynamics of our signals as well as their dependence on the terahertz field amplitude and magnetic-domain structure of our samples. In particular, the torkance inferred by comparison with a spin-dynamics model agrees well with ab initio predictions. The maximum deflection of the Néel vector ***L*** currently amounts to as much as 30°, showing that coherent ultrafast switching of ***L*** by 90° without the need for heating is within reach.

Our study has profound implications for future antiferromagnetic memory applications at high speeds and minimized energy consumption and might even serve as a blueprint for similar functionalities in multiferroic materials that inherently feature a linear coupling between electric and magnetic order [49]. Finally, the magneto-optic signals are significantly stronger in single-domain films (Supplementary Note 11), making them interesting candidates for spintronic detection of terahertz electromagnetic pulses.

## Methods

**Samples.** The samples are epitaxial Mn₂Au(001) thin films (thickness of 50 nm) grown on r-cut Al$_2$O$_3$(1$\bar{1}$02) substrates (500 μm). An additional amorphous Al$_2$O$_3$ capping layer (3 nm) serves to prevent oxidation. All layers of the heterostructures are deposited by radio-frequency sputtering [50, 51].

Following growth, one sample is exposed to an intense magnetic-field pulse (peak amplitude 60 T, duration 150 ms) [24, 34, 35]. As a result, most domains align at 0° or 180° relative to the easy $u_1$ axis, perpendicular to the applied field, with a ~2 μm domain size, as indicated in Fig.1c.

For the control measurements (Supplementary Note 11), a stack Mo(20 nm)|Ta(13 nm)|Mn$_2$Au(50 nm)|Ni$_{80}$Fe$_{20}$(10 nm)|SiN(2 nm) is deposited on a MgO(500 μm) substrate. The Py (permalloy Ni$_{80}$Fe$_{20}$) cap layer is exchange-coupled to the Mn$_2$Au layer and permits control of the antiferromagnet [29].

**Ultrafast setup.** In our terahertz-pump magneto-optic probe setup (Fig. 1a) the terahertz pump and optical probe beam are both normally incident onto the sample with spot diameters of, respectively, 950 μm and 30 μm (FWHM of the intensity). The sample azimuth $\varphi_s$ is given by the angle between the $u_1 = [110]/\sqrt{2}$ direction and the laboratory $u_x$ axis. To drive both linear and nonlinear spin dynamics, we make use of intense terahertz pump pulses (center frequency 1 THz, bandwidth ~1 THz, field strength up to 600 kV/cm, see Fig. S2). The field strength and polarity are controlled using two polyethylene wire-grid polarizers (Figs. S1, S2). The pump pulse is incident onto the metal side of the sample, and inside the Mn$_2$Au film, its electric-field amplitude amounts to about 6% of the incident amplitude (Supplementary Note 5).

To monitor the pump-induced spin dynamics, we use optical probe pulses (center wavelength 800 nm, duration 20 fs, energy 0.9 nJ) from the Ti:Sapphire oscillator seeding the amplified laser system [36]. They are linearly polarized with an angle $\varphi_{\text{pr}}$ relative to $u_x$ (Fig. 1a). Pump-induced polarization changes are detected in a balanced detection scheme, resulting in both polarization rotation and ellipticity signals $\Delta S(t)$ as a function of the pump-probe delay $t$. Note that the birefringent Al$_2$O$_3$(1$\bar{1}$02) substrate acts as a phase retardation plate when using a probe polarization that is not along the symmetry axes of the substrate. In this case, we minimize the static birefringence contribution of the substrate by adding an identical yet 90°-rotated Al$_2$O$_3$ substrate behind it.

To identify signals related to the magnetic order of Mn$_2$Au, we measure $\Delta S(t)$ for both the local Néel vector $L_0$ and its reversed version $-L_0$. While reversal of $L_0$ is generally difficult, the symmetry of Mn$_2$Au permits a simple solution: Rotation of the sample by 180° ($\varphi_s \to \varphi_s + 180°$) about the surface normal leaves the crystal structure invariant, but reverses $L_0$. The sample rotation axis coincides with the optical axis of pump and probe beams within the size of the probe focus diameter and an angle of < 5°. Consequently, the probed Mn$_2$Au volume remains the same.

**Probe signal.** In our experiment, the linearly polarized optical probe is normally incident onto the Mn$_2$Au thin film (Fig. 1a) and induces a charge-current density $j_{\text{pr}}$ that is to linear order in the probe electric field $E_{\text{pr}}$,

$$j_{\text{pr}} = \underline{\sigma} E_{\text{pr}}. \tag{5}$$

Here, $\underline{\sigma}$ is the complex-valued second-rank conductivity tensor at the frequency $\omega/2\pi$ of the probe field. The current density $j_{\text{pr}}$ emits a light field that is superimposed on the incident probe field and propagates to the polarimetric detection, which yields a signal

$$S \propto \left(E_{\text{pr}0}^* \times j_{\text{pr}}\right) \cdot u_z. \tag{6}$$

Here, $u_z$ is the sample-normal unit vector, and $E_{\text{pr}0}$ is the linearly polarized probe field in the absence of magnetic order and additional external fields. In this case, the sample is optically isotropic in the film plane, resulting in $j_{\text{pr}} \parallel E_{\text{pr}0}$ and, thus, $S = 0$. Anisotropies may lead to a nonzero signal. Note that, for the sake of simplicity, integration over all frequencies of the probe pulse and all positions of the probed volume was omitted.

It is useful to consider two basis sets: the laboratory basis $(u_x, u_y, u_z)$ and the sample-fixed basis $(u_1, u_2, u_3)$, where $u_1$, $u_2$ and $u_3$ are parallel to the crystallographic axes [110], [$\bar{1}$10] and [001] of

Mn₂Au (Fig. 1a). In our experiment, we have $u_3 = u_z$, and the probe field is in-plane (Fig. 1a). Consequently, we can write $E_{\text{pr}0} = E_{\text{pr}1} u_1 + E_{\text{pr}2} u_2$ and substitute into Eq. (6). As our probe pulse is linearly polarized, we have $E_{\text{pr}1}^* E_{\text{pr}2} = E_{\text{pr}1} E_{\text{pr}2}^*$ and, thus, obtain

$$S \propto (\sigma_{22} - \sigma_{11}) E_{\text{pr}1}^* E_{\text{pr}2} + \sigma_s \left( |E_{\text{pr}2}|^2 - |E_{\text{pr}2}|^2 \right) + \sigma_a |E_{\text{pr}0}|^2. \tag{7}$$

Here, the conductivity tensor elements are $\sigma_{ij} = u_i \cdot (\underline{\sigma} u_j)$, and the symmetric and antisymmetric off-diagonal elements are, respectively, given by

$$\sigma_s = \frac{\sigma_{21} + \sigma_{12}}{2}, \quad \sigma_a = \frac{\sigma_{21} - \sigma_{12}}{2}. \tag{8}$$

Eq. (7) shows that $\sigma_a$ induces a signal independent of the probe polarization (circular birefringence), whereas the contributions of $\sigma_s$ and $\sigma_{22} - \sigma_{11}$ depend on the probe polarization direction (linear birefringence). Due to dissipation, the $\sigma_{ij}$ are generally complex-valued and manifest themselves in both rotation and ellipticity of polarization.

To connect $S$ to the experimentally accessible sample azimuthal angle $\varphi_s = \measuredangle(u_1, u_x)$ and probe polarization angle $\varphi_{\text{pr}} = \measuredangle(E_{\text{pr}0}, u_x)$ (see Fig. 1a), we express $E_{\text{pr}1}$ and $E_{\text{pr}2}$ as $|E_{\text{pr}0}| \cos(\varphi_{\text{pr}} - \varphi_s)$ and $|E_{\text{pr}0}| \sin(\varphi_{\text{pr}} - \varphi_s)$, respectively. As a result, Eq. (7) turns into

$$S \propto |E_{\text{pr}0}|^2 [(\sigma_{22} - \sigma_{11}) \sin(2\varphi_{\text{pr}} - 2\varphi_s) + \sigma_s \cos(2\varphi_{\text{pr}} - 2\varphi_s) + \sigma_a]. \tag{9}$$

The three terms contributing to the signal in Eq. (9) can be experimentally separated based on their different dependence on the angle $\varphi_{\text{pr}} - \varphi_s = \measuredangle(E_{\text{pr}0}, u_1)$ between probe field and Mn₂Au crystal axis $u_1$, thereby yielding $\sigma_{22} - \sigma_{11}$, $\sigma_s$ and $\sigma_a$.

To relate $S$ to the magnetic order of the sample and the impact of the terahertz pump field $(E, B)$, one can expand the $\sigma_{ij}$ up to linear order in the pump electromagnetic field $(E, B)$ and up to second order in the magnetization $M$ and Néel vector $L$. The in-plane projection $L_\parallel = L - (u_z \cdot L) u_z$ of $L$ is written as $L_\parallel = L_\parallel \cos \varphi_L u_1 + L_\parallel \sin \varphi_L u_2$ (see Fig. 1a), and the equilibrium values of $M$ and $L$ are $M_0 = 0$ and $L_0$. The spatial symmetries of our Mn₂Au sample and the normally incident pump pulse greatly simplify the equations, resulting in

$$\sigma_{22} - \sigma_{11} = a L_\parallel^2(t) \cos[2\varphi_L(t)] + \eta_\mathcal{N} L_{0\parallel} E(t) \sin(\varphi_E - \varphi_s + \varphi_{L_0}),$$
$$\sigma_s = b L_\parallel^2(t) \sin[2\varphi_L(t)] + \kappa_\mathcal{N} L_{0\parallel} E(t) \cos(\varphi_E - \varphi_s + \varphi_{L_0}), \tag{10}$$
$$\sigma_a = c M_z(t) + \mu_\mathcal{N} L_{0\parallel} E(t) \cos(\varphi_E - \varphi_s - \varphi_{L_0}).$$

The coefficients $a$, $b$, $c$, $\eta_\mathcal{N}$, $\kappa_\mathcal{N}$ and $\mu_\mathcal{N}$ are independent of $M$, $L$, $E$ and $B$. In our experiment, $\varphi_{L_0}$ is an integer multiple of 90°.

The first term in each Eq. (10) monitors true spin dynamics $M(t)$ and $L(t)$. In contrast, the second term in each Eq. (10) arises from the dynamics of pump-induced changes in the non-spin degrees of freedom $\mathcal{N}$ of Mn₂Au. As shown in Supplementary Note 12 for each Eq. (10), the pump-induced changes in the first term have the same dependence on $\varphi_E$, $\varphi_{L_0}$ and $\varphi_s$ as the second term. Consequently, signals due to true spin dynamics and the dynamics of non-spin degrees of freedom cannot be separated based on the variation of $\varphi_E$, $\varphi_{L_0}$ and $\varphi_s$. Therefore, complementary arguments need to be provided (see Supplementary Note 4).

Note that each second term on the right-hand side of Eq. (10) is odd in both $E$ and $L_0$. One can generalize this property by symmetry arguments (see Supplementary Note 12). It follows that, in general, any signal contribution odd in $E$ is simultaneously odd in $L_0$ and vice versa.

We finally combine of Eqs. (9) and (10) to obtain

$$\Delta S \propto a \sin(2\varphi_{\text{pr}} - 2\varphi_s) \Delta(L_\parallel^2 \cos(2\varphi_L)) + b \cos(2\varphi_{\text{pr}} - 2\varphi_s) \Delta(L_\parallel^2 \sin(2\varphi_L)) + c u_z \cdot \Delta M + \Delta S_\mathcal{N}, \tag{11}$$

where Δ indicates pump-induced changes.

**Spatial averaging of the probe.** The probe signal is an integral over the probed volume. As the domains of our sample (~2 μm) are much smaller than the probe spot (≈30 μm), we expect averaging over the magnetic domain configuration of the given sample. To determine the signal from a multi-domain sample, we apply spatial averaging $\langle . \rangle$ to Eq. (11) and obtain Eq. (2) of the main text. As detailed there, the observed signal symmetry allows us to write the dominant second term of Eq. (2) in its linearized form as

$$\Delta S(t) \propto \langle \Delta(\boldsymbol{L}_{\|0}^2) \sin(2\varphi_{L_0}) \rangle + 2\langle \boldsymbol{L}_{\|0}^2 \cos(2\varphi_{L_0}) \Delta\varphi_L \rangle. \tag{12}$$

In the following, we focus on the prealigned sample, where $\varphi_{L_0} = 0°$ and 180° [24]. Therefore, $\sin(2\varphi_{L_0}) = 0$, $\cos(2\varphi_{L_0}) = 1$, and $\boldsymbol{L}_{\|0}^2$ is the same for the 0° and 180° configurations, and Eq. (12) turns into Eq. (3) of the main text.

To consider the spatial averages in Eq. (2) beyond the linear limit, we note that signals $\propto \Delta(\boldsymbol{L}_{\|}^2)$ are negligible in the linearized case (see Eq. (12)). We assume the same for larger signals, i.e., $\boldsymbol{L}_{\|}^2 \approx \boldsymbol{L}_{0\|}^2$, and obtain

$$\begin{aligned}\Delta S &\propto a \sin(2\varphi_{\mathrm{pr}} - 2\varphi_{\mathrm{s}}) \langle \boldsymbol{L}_{0\|}^2 [\cos(2\varphi_{L_0} + 2\Delta\varphi_L) - \cos(2\varphi_{L_0})] \rangle \\ &+ b \cos(2\varphi_{\mathrm{pr}} - 2\varphi_{\mathrm{s}}) \langle \boldsymbol{L}_{0\|}^2 \sin(2\varphi_{L_0} + 2\Delta\varphi_L) \rangle + c \boldsymbol{u}_z \cdot \langle \Delta\boldsymbol{M} \rangle,\end{aligned} \tag{13}$$

where $\langle \cdot \rangle$ denotes the average over the probed volume. As $\varphi_{L_0} = 0°$ or 180°, Eq. (13) simplifies to

$$\begin{aligned}\Delta S &\propto a \sin(2\varphi_{\mathrm{pr}} - 2\varphi_{\mathrm{s}}) \boldsymbol{L}_{0\|}^2 \langle \cos(2\Delta\varphi_L) - 1 \rangle + b \cos(2\varphi_{\mathrm{pr}} - 2\varphi_{\mathrm{s}}) \boldsymbol{L}_{0\|}^2 \langle \sin(2\Delta\varphi_L) \rangle \\ &+ c \boldsymbol{u}_z \cdot \langle \Delta\boldsymbol{M} \rangle,\end{aligned} \tag{14}$$

where the average $\langle . \rangle$ runs over just $\varphi_{L_0} = 0°$ and 180°. For small deflections $\Delta\varphi_L$, Eq. (14) can be expanded up to quadratic order as

$$\Delta S \propto 2a \sin(2\varphi_{\mathrm{pr}} - 2\varphi_s) \boldsymbol{L}_{0\|}^2 \langle \Delta\varphi_L^2 \rangle + 2b \cos(2\varphi_{\mathrm{pr}} - 2\varphi_s) \boldsymbol{L}_{0\|}^2 \langle \Delta\varphi_L \rangle + c \boldsymbol{u}_z \cdot \langle \Delta\boldsymbol{M} \rangle. \tag{15}$$

The deflection angle $\Delta\varphi_L$ and out-of-plane magnetization $\Delta M_z = \boldsymbol{u}_z \cdot \Delta\boldsymbol{M}$ are odd in both $\boldsymbol{E}$ and $\boldsymbol{L}_0$, as $\Delta S$ is required to be odd in $\boldsymbol{E}$ and $\boldsymbol{L}_0$ by definition. It follows that $\Delta\varphi_L^{0°} = -\Delta\varphi_L^{180°}$ and $\Delta M_z^{0°} = -\Delta M_z^{180°}$. Assuming the $\boldsymbol{L}_0$-dependence of $\Delta\varphi_L^{0°}$ and $\Delta M_z^{0°}$ is linear, the spatial average of both $\Delta\varphi_L$ and $\Delta M_z$ scales with $\langle \boldsymbol{L}_0 \rangle$ in the prealigned sample. Therefore, observation of a signal from the second term of Eq. (15) requires $\langle \boldsymbol{L}_0 \rangle \neq 0$. This contribution has a smaller amplitude than in the case of a single domain, but unaltered dynamics.

The preceding argumentation also holds for the $\langle \sin(2\Delta\varphi_L) \rangle$ term in Eq. (14) and, thus, includes nonlinear responses odd in both $\boldsymbol{E}$ and $\boldsymbol{L}_0$.

The origin of $\langle \boldsymbol{L}_0 \rangle \neq 0$ lies in the domain reorientation by the strong magnetic-field pulse, as discussed in Supplementary Note 5.

**Micromagnetic model details.** Two uniform antiferromagnetic magnon modes are predicted to exist in Mn$_2$Au. They exhibit predominantly in-plane and out-of-plane oscillation character of the Néel vector, respectively [40]. The dynamics of the in-plane mode is captured by the uniform azimuthal rotation $\varphi_L(t)$. The equation of motion for this mode in a given domain driven by NSOTs is [40, 42]

$$\ddot{\varphi}_L + 2\gamma \alpha_G B_{\mathrm{ex}} \dot{\varphi}_L + (\Omega_0^2/4) \sin(4\varphi_L) = -\gamma B_{\mathrm{ex}} \lambda_{\mathrm{NSOT}} \sigma [(\boldsymbol{E} \cdot \boldsymbol{u}_1) \cos\varphi_L + (\boldsymbol{E} \cdot \boldsymbol{u}_2) \sin\varphi_L]. \tag{16}$$

Here, $\gamma$ is the gyromagnetic ratio, $B_{\mathrm{ex}} \approx 1300$ T is the inter-sublattice exchange field [31], $\alpha_G$ is the Gilbert damping parameter, $\Omega_0^2 = 4\gamma^2 B_{\mathrm{ex}} B_{\mathrm{ani}}$, where $B_{\mathrm{ani}}$ is the in-plane anisotropy field in the definition of Ref. [40], $\lambda_{\mathrm{NSOT}}$ is the NSOT coupling strength (or torkance), and $\sigma \approx 1.5$ MS/m is the approximate measured conductivity at 0-3.5 THz. The terahertz electric field $\boldsymbol{E}(t)$ inside the thin film amounts to about 6% of the incident field (Figs. S6, S7).

Eq. (16) implies that the in-plane AFM mode frequency is fully determined by the in-plane anisotropy and exchange fields. The spin-flop field is connected to these quantities by $B_{\mathrm{sf}} = 2\sqrt{B_{\mathrm{ani}} B_{\mathrm{ex}}}$ [40] and, thus, to the bare magnon frequency $\Omega_0/2\pi$ through $\Omega_0 \sim \gamma B_{\mathrm{sf}}$. This relationship allows us to directly relate the measured $\Omega_0$ to results from complementary experiments [24, 35].

When going from one domain with equilibrium orientation $\varphi_{L_0}$ to the oppositely oriented domain with angle $\varphi_{L_0} + 180°$, the right-hand side of Eq. (16) changes sign. Therefore, the direction of rotation of $\boldsymbol{L}$ also changes sign, from, e.g., clockwise to counterclockwise (Fig. 1c).

The second magnon mode involves an out-of-plane oscillation of $\boldsymbol{L}$ and is determined by the strong uniaxial out-of-plane anisotropy field $|B_{\mathrm{ani}}^{\parallel}| \geq 5$ T [32]. It is not considered here because of the following reasons. First, its symmetry only allows excitation by the terahertz magnetic field. The corresponding signal would be even in $\boldsymbol{L}_0$. Second, the only detectable signal component in our experimental geometry would be $\propto \left(\boldsymbol{u}_y \cdot \Delta \boldsymbol{M}\right)^2$, which is very small compared to the signal due to deflection $\Delta \boldsymbol{L}$ of the Néel vector.

**Acknowledgments**

We acknowledge funding by the German Research Foundation through the collaborative research centers SFB TRR 227 "Ultrafast spin dynamics" (project ID 328545488, projects A05 and B02), SFB TRR 173 "Spin+X" (project ID 268565370) and priority program SPP2314 INTEREST (project ITISA), the European Union through the projects ERC H2020 CoG TERAMAG/grant no. 681917 and ERC SyG 3D MAGiC/grant no. 856538. We acknowledge KAUST (Grant No. OSR-2019-CRG8-4048) and support by the European Commission under FET-Open Grant Agreements No. 863155 (s-Nebula) and No. 766566 (ASPIN).

OGomonay acknowledges support from the Deutsche Forschungsgemeinschaft (DFG, German Research Foundation) - TRR 173 – 268565370 (project A11 and B12), and also ASPIN: EU FET Open RIA Grant no. 766566.

**Author contributions**

TK, YB, ALC conceived the experiment. YB, ALC, OGueckstock performed the experiments. YB, ALC, OGueckstock analyzed the data. SYB prepared samples. SYB, SR, MJ characterized samples. OGomonay provided the basis of the theoretical model. YB, ALC and TK wrote the original draft. TSS, MW, OGomonay, MK, SR and MJ reviewed and edited the manuscript. All authors read and commented on the manuscript.

**Competing interests**

The authors declare that they have no competing interests.